\documentstyle[epsfig]{mn}

\def\ho{$H_{0}$}
\def\sz{SZ} 
\def\xray{X-ray}
\def\microJy{$\mu$Jy}
\def\milliJy{mJy}
\def\kl{k$\lambda$}
\def\ra#1h#2m#3.#4s{$\rm {RA} \ #1^h \ #2^m \ #3^s#4 \ $}
\def\dec#1d#2m#3s{$\rm {Dec.}\ +#1^{\circ} \ #2' \ #3''$}
\def\msolar{$M_{\odot}$}
\def\hounit{km s$^{-1}$ Mpc$^{-1}$} 

\def\arcsec{\hbox{$^{\prime\prime}$}}
\def\asca{\textit{ASCA}}
\def\rosat{\textit{ROSAT}}

\def\clean{\textsc{Clean}}
\def\eg{e.g.\ }
\def\ie{i.e.\ }

\def\aj{{AJ}}                   
\def\apjs{{ApJS}}               
\def\mnras{{MNRAS}}             
\def\apj{{ApJ}}                 
\def\apjl{{ApJ}}                

\newenvironment{centre}
 {\begin{center}} 
 {\end{center}} 

\title[Removal of radio sources in SZ observations]{A maximum-likelihood approach to
removing radio sources from SZ observations, with application to Abell 611}

\author[W.~F.~Grainger et al.]{William F. Grainger, Rhiju
Das\thanks{Present address: Physics Department, Stanford University, CA, 94305-4060, USA.}, Keith Grainge, Michael E. Jones, \cr R\"udiger
Kneissl, G.\ G.\ Pooley, Richard Saunders \\ 
Astrophysics, Cavendish Laboratory, Madingley Road, Cambridge CB3 0HE, UK.}
\begin{document}

\maketitle

\begin{abstract}
We describe a maximum-likelihood technique for the removal of contaminating radio
sources from interferometric observations of the Sunyaev-Zel'dovich
(\sz) effect. This technique, based on a simultaneous
fit for the radio sources and extended \sz\
emission, is also compared to techniques previously applied to Ryle
Telescope observations and is found to be robust. The technique is
then applied to new observations of the cluster Abell 611, and a
decrement of $-540 \pm 125$~\microJy\ beam$^{-1}$ is found. This is
combined with a \rosat\ HRI image and a published \asca\ temperature to give an \ho\ estimate of
 $52^{+24}_{-16}$~\hounit.
\end{abstract} 

\section{Introduction}
This paper is concerned with the subtraction of radio sources which
would otherwise contaminate or obliterate detections of the
Sunyaev-Zel'dovich (\sz) effect \cite{sz1,sz2} towards galaxy
clusters. The work described here is in connection with the Ryle
Telescope (RT, see \eg Jones et al.\ 2001 and Grainge et
al.\ 2001b), but the issues are relevant to all
cm-wavelength \sz\ observations with interferometers (see \eg Reese et
al.\ 2000). For a massive cluster at moderate or
high redshift, the flux that the RT detects from the \sz\ effect at
15~GHz is typically $-$500~\microJy\ on its shortest baselines. This
is sufficiently faint that radio sources will almost invariably be
present with comparable or greater amplitudes. Thus removing the
effects of radio sources is an essential step. We describe and compare
two past methods of measuring \sz\ decrements in the presence of
sources as well as a maximum-likelihood method. We then apply this to new RT
observations of cluster A611 which we combine with \xray\ data to
estimate \ho. All coordinates are J2000 and, except where otherwise
stated, we use an Einstein-de-Sitter world model.

\section{Removing radio sources from \sz\ observations with the RT}

\begin{figure}
\begin{centre}
\epsfig{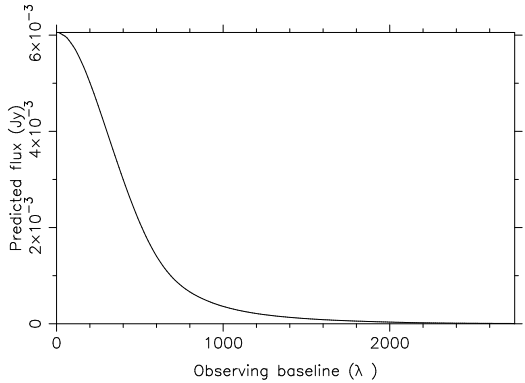}
\caption{The flux density as a function of baseline for a spherical King-model
cluster with central electron density $=10^4 {\rm m}^{-3}, \beta=0.65,
\mbox{core radius} = 60\arcsec$, $T_e=7.8\times10^{7}~\mbox{K}$,
$z=0.171$ and \ho=50 \hounit, as observed
with the RT. (From Grainge 1996). The shortest RT
physical baselines are 870~$\lambda$; this becomes somewhat shorter
with projection.}
\label{fig:cluster}
\end{centre}
\end{figure}

As the RT is an interferometer with a wide range of baselines, it can
simultaneously measure the extended \sz\ flux and the fluxes and
positions of the small angular size radio
sources. Figure~\ref{fig:cluster} illustrates the variation of \sz\
flux with baseline for the RT when observing a massive Abell cluster
at $z=0.171$. The variation with redshift is slight over the range
$z=0.15$--$5$, as shown in Grainge et al.\ \shortcite{1413}. The \sz\
effect has effectively been completely resolved out for baselines
above $\simeq 1.5$~\kl, and so these ``long'' baselines can be used to
measure sources and so remove their effects from the \sz\ signal seen
on the ``short'' baselines. The measurements are simultaneous and, of
course, at the same frequency, and so the spectral index is
unimportant. Variability is unimportant if the telescope configuration
does not change; see Grainge et al.\ \shortcite{1413RT} for details.  By
choosing an interferometer configuration such that there are more long
baselines than short, it is possible to optimise the observations to
achieve good signal-to-noise for the \sz\ effect without it being
dominated by noise from unsubtracted sources.

There are three methods of source subtraction that have been applied
to RT data:

\begin{enumerate}
\item
using \textsc{Clean}, which has been used for the bulk of
published \sz\ measurements from the RT (\eg Grainge et al.\ 1996);
\item
the matrix method, which was first used by Grainge \shortcite{keiththesis}; and
\item
the \textsc{FluxFitter} method, first used by Das
\shortcite{rhijuthesis}, and used here in modified form.
\end{enumerate}

This sequence began with \textsc{Clean}, the classical radio-astronomy
image deconvolution technique (see \eg Greisen 1994 and
Perley et~al.\ 1989). The matrix method is a  linear
method for removing the effects of sidelobes, and \textsc{FluxFitter}
-- using maximum likelihood -- addresses the problem of simultaneously
fitting both radio sources and \sz\ decrement.

\subsection{The test data}
The three methods can be  explained and compared with an
example of a simulated dataset containing both sources and an \sz\
effect. The simulation is of a 54 $\times$ 12-hour long observation of
a field at declination $44^{\circ}$.  The $uv$ coverage is based on a
standard Cb configuration for the RT; in this configuration four
aerials are parked on an east-west rail-track at locations 36, 72, 90
and 108~m from the closest fixed aerial (see Grainge et al.\
1996 for details).  The point-source fluxes and positions
are shown in Table~\ref{tab:modelfluxes}, and the noise level was set
to 7~\milliJy~visibility$^{-1/2}$, corresponding to
200~\microJy~day$^{-1/2}$, as expected for a standard RT observation
with five aerials. The \sz\ decrement is
based on Abell 2218, and is modelled as an isothermal ellipsoid with
a King electron density profile \cite{king} at the centre of the map
with a central electron density of $10^{4} {\rm m}^{-3}$, $\beta =
0.65$, a temperature of $8.7\times10^{7}$~K
and core radii of 60\arcsec\ and 40\arcsec\ on the sky and 49\arcsec\ ($=\sqrt{60\times40}$)
along the line of sight. The central temperature decrement for this
cluster is 0.82~mK and, as observed with the RT, the cluster gives
$-$660~\microJy\ beam$^{-1}$ on the shortest baseline.

\begin{table}
\caption{The source fluxes and positions in the simulated data
set. The convention used in this and other tables is: positive
$\Delta$RA is an increase in the RA value, so the source is to the
East (to the left on conventional maps) and positive $\Delta$Dec is an
increase in Dec, i.e. to the North.}
\label{tab:modelfluxes}
\begin{tabular}{lll}
Source number &Flux & Offset from pointing centre \\
               &(\microJy\ beam$^{-1}$) & (arcseconds) \\
  1  &    2960   &   $-$10, 10  \\
  2  &    910    &    35,  15 \\
  3  &    255    &   $-$50, $-$40 \\ 
  4  &    170    &  $-$120, 100 \\
  5  &    100    &     0,   0 \\
  6  &     80    &    60, $-$60 \\
\end{tabular}
\end{table}

\subsection[The \clean\ method]{The C{\footnotesize LEAN} method}

For this method, a dirty (\ie un\clean ed) map of all the baselines
longer than 1.5~\kl\ is produced within \textsc{Aips} \cite{aips}.
Natural weighting is used to give the best possible signal-to-noise
ratio at the expense of larger sidelobes.  The map is \clean ed in the
standard way by placing \clean\ boxes around the obvious sources.
After deconvolution, the source fluxes are measured, using the
\textsc{Aips} verb \textsc{Maxfit}, which interpolates the position
and value of the maximum flux density. The measured fluxes are then
removed from the visibility data using the task \textsc{Uvsub}.

\begin{figure}
\begin{centre}
\epsfig{file=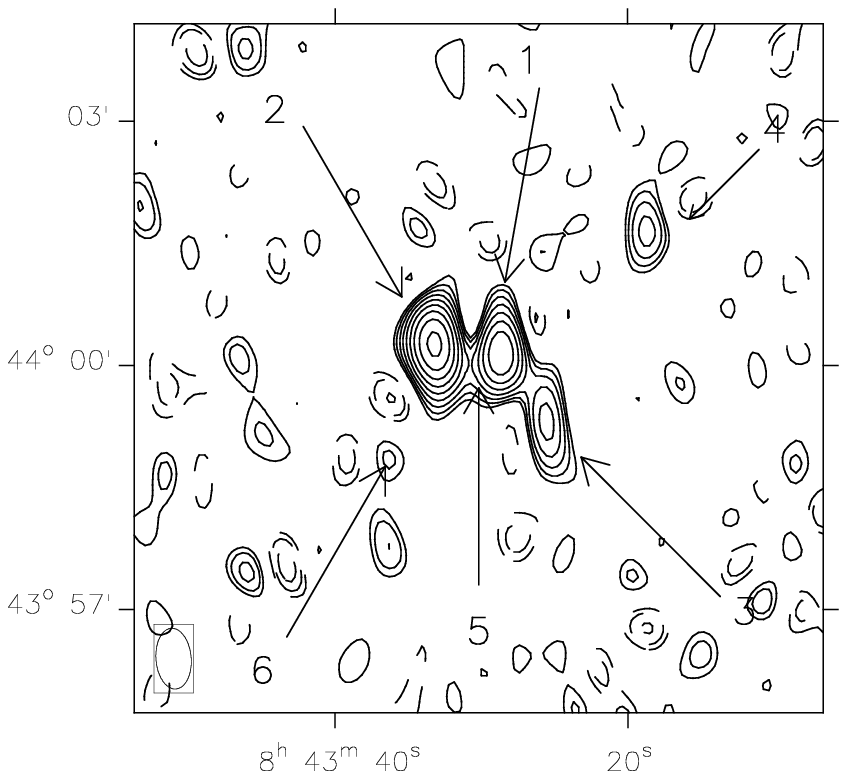}
\caption{A \textsc{Clean}ed map of the simulated data made with
baselines longer than 1.5~\kl.  Source 1 has had 2500\microJy\
beam$^{-1}$\ of flux already removed to aid identification of weaker
sources. The contour levels are $-$55$\times$(1, $\surd{2}$, 2)~\microJy\ beam$^{-1}$\ (dashed),
55$\times$(1, $\surd{2}$, 2, $\surd{8}$, 4,~\ldots)~\microJy\
beam$^{-1}$ (solid). The beamsize FWHM (45\arcsec\ $\times$ 26\arcsec\
at 4$^{\circ}$) is shown in the bottom left. The noise in the map due
to the system temperature is 38~\microJy~beam$^{-1}$.}
\label{fig:cleaned}
\end{centre}
\end{figure}

The sources and the positions found in the simulated dataset are
listed in Table~\ref{tab:cleanpositions}, and the \textsc{Clean}ed map
is shown in Figure~\ref{fig:cleaned}. The noise in the map due to the
system temperature is 38~\microJy\ beam$^{-1}$. The six model sources
are labelled; note that only four brightest were found.

A map made with baselines longer than 1.5~\kl\ was consistent with noise after these four
sources had been subtracted. The \sz\ flux observed was
$-675 \pm 72$~\milliJy\ beam$^{-1}$, at a position ($-4$, $-2$) arcsec from the pointing
centre. The \sz\ values were measured with \textsc{Maxfit} from a map
made with the baselines shorter than 1.0~\kl.

\begin{table}
\caption{Fluxes and positions as measured by \textsc{Maxfit} after 
\clean ing with four \clean\ boxes. The noise on each flux is 38~\microJy. }
\label{tab:cleanpositions}
\begin{tabular}{lll}
Source number & Flux     & Offset from pointing centre \\
              & (\microJy\ beam$^{-1}$) &    (arcseconds)               \\
1 & 3040 &   $-$9.6, 9.6     \\
2 & 970  &     34.5, 15.8    \\
3 & 270  &  $-$49.9, $-$45.8 \\
4 & 170  & $-$120.1, 91.9    \\
\end{tabular}
\end{table}

\begin{table}
\caption{Fluxes and positions from the \clean\ method with six \clean\
boxes.}
\label{tab:clean2}
\begin{tabular}{lll}
Source number & Flux     & Offset from pointing centre \\
              & (\microJy\ beam$^{-1}$) &    (arcseconds)               \\
1 & 3020 &   $-$9.6, 9.5     \\
2 & 950  &     34.5, 15.8    \\
3 & 240  &  $-$47.6, $-$42.0 \\
4 & 180  & $-$121.6, 98.4    \\
5 & 85   & 12,$-$7 \\
6 & 82   & 70,$-$70 \\
\end{tabular}
\end{table}

A comparison of the sources found (Table~\ref{tab:cleanpositions}) and
the sources actually in the model (Table~\ref{tab:modelfluxes}) shows
the how well the \clean\ method works. The most glaring problem is
that only four out of the original six sources have been detected.  As
lower-frequency surveys such as FIRST,  NVSS or optical images often provide
information about the source envirnoment at 15~GHz, the long-baseline
map is rarely used in isolation. As such, a second test was performed,
placing the \clean\ boxes as before, but adding small \clean\ boxes
around the locations of the other two weaker sources. The positions
and fluxes of the six sources are shown in Table~\ref{tab:clean2}. The
noise on the long-baseline map is 38~\microJy\ beam$^{-1}$. Thus
additional radio or optical information leads to the detection of the
two faintest sources.

\subsubsection{Source finding}

Source finding is  done in a non-linear, iterative manner. A
typical source-subtraction process may involve many iterations of map
making, running the subtraction algorithm on the sources found, mapping
residuals, finding another source and then adding that into the model.

A high signal-to-noise source is easy to identify, and causes no
problems. However, the process becomes subjective around
signal-to-noise ratios of 4 to 3.5. This is the point at which the
non-Gaussian, correlated statistics in the image plane conspire with
the high RT sidelobes (when using a few antennas for source finding)
to make source identification more difficult. 

Radio sources in the clusters we observe have angular sizes
smaller than the maximum resolution used.  However, the number of
\clean\ components is generally much larger than the number of sources
in the field. With the test data, only four sources were identified,
but \textsc{Aips} produced a model with 28 \clean\ components. Since
the real sources are evidently point sources, this is over-modelling
the data; and potentially  biasing.  There is also a degree of
subjectivity in the placing and  sizing of \textsc{Clean} boxes. 

\subsection{The matrix method}

The matrix method was initally developed by KG \cite{keiththesis}, and
we here describe it and assess its performance.  In the matrix
method, sources are first identified from a un\textsc{Clean}ed
long-baseline map, and \textsc{Maxfit} used to measure the positions
and the fluxes of the sources. The convolution that occurs when
observing with an interferometer means that the flux on the dirty map
of the $j^{\rm th}$ source, $S_{\mathrm{dirty},j}$, is given by
\begin{equation}
 S_{\mathrm{dirty},j} = \sum_{i=1}^{n} S_{{\rm sky},i}\,B_{i,j}\,P_{i},
\end{equation}
where $S_{{\rm sky},i}$ is the true flux on the sky of the $i^{th}$ point
source, $B_{i,j}$ is a factor due to the synthesised beam that depends
on the displacement (on the sky) between the source in question and
the $i^{th}$ source, and $P_{i}$ is the primary (envelope) beam
attenuation. The value of $B_{i,j}$ is directly measured from the
dirty beam produced by \textsc{Horus}. 
There is thus a matrix equation linking the measured dirty fluxes with
the true sky fluxes, which is solved by inverting the matrix.
This method has the advantage over \textsc{Clean} in that
sources that are measured in the map to be point sources are modelled
with one flux and position. With the simulated test data, the four
sources identified with the \textsc{Clean} method were used. The
resulting matrix was
\begin{equation}
\left(
\begin{array}{c} 2875 \\ 473 \\ 105 \\ 585 \\
\end{array}
\right)
=
\left(
\begin{array}{cccc}
 1     &  0.13   &  -0.03  &  0.16  \\
-0.13  &  1      &   0.02  & -0.04  \\
-0.03  &  0.02   &   1     &  0.12  \\
 0.16  & -0.04   &   0.12  &  1     \\
\end{array}
\right)
 \left(\begin{array}{c} S_{{\rm sky},1} \\
S_{\mathrm{sky},2} \\
S_{\mathrm{sky},3} \\
S_{\mathrm{sky},4} \\
\end{array}
\right),
\end{equation}
where the vector on the left side of the equation is a measurement of
the dirty flux (in \microJy\  beam$^{-1}$) at each point. The value
$S_{{\rm sky},x}$ is the beam-attenuated flux on the sky of source
$x$, using the same labelling as for sources found in the \textsc{Clean}
method. 

The matrix method is linear, an apparent advantage over the \clean\
method. If after the first subtraction attempt some sources are still
present, then the additional terms for the matrix can then be measured
and the solution recalculated. 

\begin{figure}
\begin{centre}
\epsfig{file=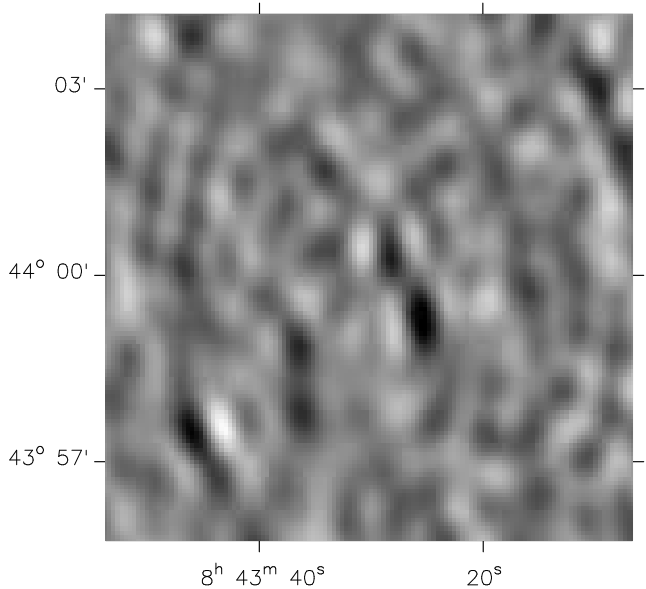}
\caption{Long -- i.e.\ greater than 1.5~\kl\ -- baseline map for the test
data after the sources in Table~\ref{tab:Matrixpositions} have been
subtracted. The greyscale range is from -200 (light) to 200 \microJy\
beam$^{-1}$.}
\label{fig:matrixsubbed}
\end{centre}
\end{figure}

Table~\ref{tab:Matrixpositions} shows the fluxes found. A comparison
with the sources known to be in the model
(Table~\ref{tab:modelfluxes}) shows that the flux of source three is
significantly different. There are two reasons for this.  Firstly, not
all the flux has been subtracted -- see Figure~\ref{fig:matrixsubbed}.
Secondly, the positions used have been determined from the
un\textsc{Clean}ed map, by searching for extrema with
\textsc{Maxfit}. Occasionally, the relative positions of sources are
such that one source is in the steepest part of a sidelobe of another,
and \textsc{Maxfit} does not find an extremum that corresponds to that
source. This has occurred in this case with source 4. In this
situation, \textsc{Maxfit} is not used, and the position from the
\textsc{Clean} method is used. The flux value from the
un\textsc{Clean}ed map in that position is then used.  The reliance on
the \clean\ method combined positional problems and with
difficulties automating the process for large numbers of sources means
that the ``matrix method'' has only been applied in simple cases with
a few well separated sources.

\begin{table}
\caption{Source fluxes and positions from the matrix method.}
\label{tab:Matrixpositions}
\begin{tabular}{lll}

Source number & Flux & Offset from pointing centre \\ 
  & (\microJy\  beam$^{-1}$) & (arcseconds) \\
1 & 2970  & $-$9, 9.6 \\
2 & 877 & 34, 15 \\
3 & 8 & $-$47, $-$42.8 \\
4 & 142 & $-$122, 99.9 \\
\end{tabular}
\end{table}

\subsection[The \textsc{FluxFitter} method]{The F{\footnotesize
LUX}F{\footnotesize ITTER} method}
\label{sc:positiontest}
In an attempt to overcome the problems of the \textsc{Clean} and
matrix methods, the \textsc{FluxFitter} algorithm was introduced.  The
algorithm that it uses is straightforward:

\begin{enumerate}
\item
An initial model of the sky is made, using a set of
parameters that represents the positions and fluxes of each 
source, including the \sz\ decrement. This model is determined from
the long-baseline RT maps (either raw or \clean ed) and, for the \sz\
parameters, from the \xray\ image.
\item
The flux that the RT would observe is calculated for every visibility
point.
\item
The misfit between the model and real $uv$ data is then calculated as
$\chi^2$. 
\item 
The parameters are then varied to minimise $\chi^2$.
\end{enumerate}

 The best-fitting parameter values are then used to subtract the radio
sources; this is done within \textsc{Aips}.  There are two
advantages to this method. It works almost entirely in the aperture
plane; only the source identification and approximate position finding
is done in the image plane.  Working in the aperture plane is preferred
because the noise distribution is known to be Gaussian. The second
advantage of this method is in point (i): a simultaneous fit to the
positions and fluxes of the point sources and the \sz\ decrement is a clear
improvement over either of the previous methods as there is no arbitrary
``long'' and ``short'' baseline split, and
it allows full use of all visibility data, which increases the
signal-to-noise ratio for point-source measurements.

The initial parameters for the sources are still estimated by iterating
the \textsc{Clean} method. In a complex situation with both bright and
faint sources, the map is \textsc{Clean}ed, and then positions
measured. These sources are then subtracted and the subtracted data then
mapped again. This loop can be performed many times to estimate the
number of sources and their approximate positions and fluxes. In a
less complex situation, \textsc{Clean} is not used, and sources are 
approximately subtracted and then the data remapped. Again, this is
just to provide an initial guess for \textsc{FluxFitter}.

Currently only the amplitude of the \sz\ decrement is varied. The
other parameters that describe the decrement -- position, core radius
and $\beta$-parameter for the cluster -- are all fixed in advance from
the \xray\ data.  This is done as the radio data do not
constrain well the core-radius or $\beta$ value; with present telescopes,
the \xray\ measurements constrain the core radius, $\beta$ value and
position much better.

\textsc{FluxFitter} was run twice on the simulated test data. The
fluxes and positions from the \clean\ method were used as an initial
guess for the first run.  The fluxes and positions from this run are
shown in Table~\ref{tab:fluxfitterpositions1}; the errors are those
reported by \textsc{FluxFitter} -- see below.  After subtraction with
\textsc{Uvsub}, the map of baselines greater than 1.5~\kl\ was
consistent with noise. For a second run, the fluxes and positions from
the \clean\ method were used, and the positions of the additional two
sources were also used.  The fluxes and positions from this run are
shown in Table~\ref{tab:flux2}. This table shows that the two
additional sources are detected with good significance. That the
overall noise level of 30~\microJy\ beam$^{-1}$\ is lower than that
for the \clean\ method is not surprising as all the baselines are
being used in the determination of the fluxes and positions.  After
subtracting the six reported sources, a map of baselines shorter than
1.0~\kl\ shows a decrement of flux $-700\pm65$~\milliJy\ beam$^{-1}$\
at an offset (9\arcsec,4\arcsec). \textsc{FluxFitter} itself finds a
central decrement of $0.87 \pm 0.08$~mK, close to the set value of 0.82~mK.

As an additional check, a third run of \textsc{FluxFitter} was
performed, and the four sources found from the \clean\ method and two
random points were used as the initial guess. In this case,
\textsc{FluxFitter} reported that both of the two random ``sources''
had fluxes below the noise level and very large positional-errors.

\begin{table}
\caption{Source positions and fluxes as reported by
\textsc{FluxFitter} using the \clean\ model as an initial guess. The
formal error on each flux is 30~\microJy~beam$^{-1}$.}
\label{tab:fluxfitterpositions1}
\begin{tabular}{lll}

Source number & Flux     & Offset from pointing centre \\
              & (\microJy\  beam$^{-1}$) &    (arcseconds)              \\
1 & 3031  &   $-9.5 \pm 0.1$,   $9.6 \pm 0.2$ \\
2 & 966   &   $34.5 \pm 0.3$,  $15.3 \pm 0.5$ \\
3 & 208   &  $-47.5 \pm 1.2$, $-42.8 \pm 2.8$ \\
4 & 184   & $-122.3 \pm 1.6$,  $98.2 \pm 2.8$ \\
\end{tabular}
\end{table}

\begin{table}
\caption{Source positions and fluxes as reported by
\textsc{FluxFitter} using additional information. The
formal error on each flux is 30~\microJy~beam$^{-1}$.}
\label{tab:flux2}
\begin{tabular}{lll}
Source number & Flux     & Offset from pointing centre \\
              & (\microJy\  beam$^{-1}$) &    (arcseconds)              \\
1 & 3024 &   $-9.9 \pm 0.1$,   $9.9 \pm 0.2$ \\
2 & 970  &   $34.9 \pm 0.5$,  $15.6 \pm 0.7$ \\
3 & 212  &  $-47.8 \pm 1.8$, $-40.8 \pm 3.6$ \\
4 & 155  & $-121   \pm 2.6$,  $98.8 \pm 4.1$ \\
5 & 140  &    $7.9 \pm 3.8$,  $-3.1 \pm 4.2$ \\
6 & 132  &   $67.4 \pm 3.3$, $-64.8 \pm 4.6$ \\
\end{tabular}
\end{table}

\begin{figure}
\begin{centre}
\epsfig{width=5cm,angle=-90,file=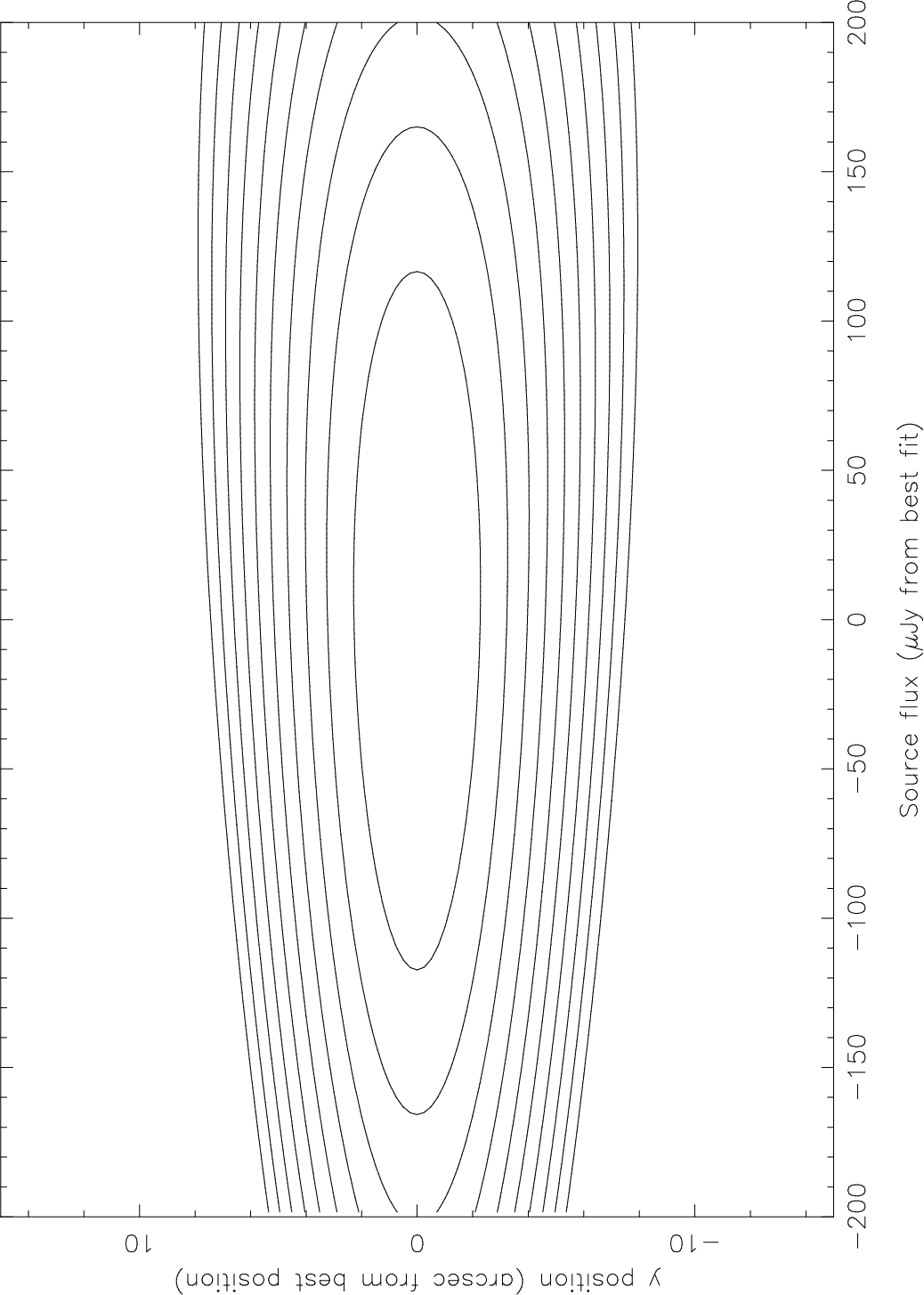}
\caption{$\chi^{2}$ contours for the flux and position in Dec for the
brightest source in the test field.The spacing between contour levels in each figure is such that the reduced $\chi^2$ value increases
by 1 between each contour.}
\label{fig:chiplots}
\end{centre}

\begin{centre}
\epsfig{width=5cm,angle=-90,file=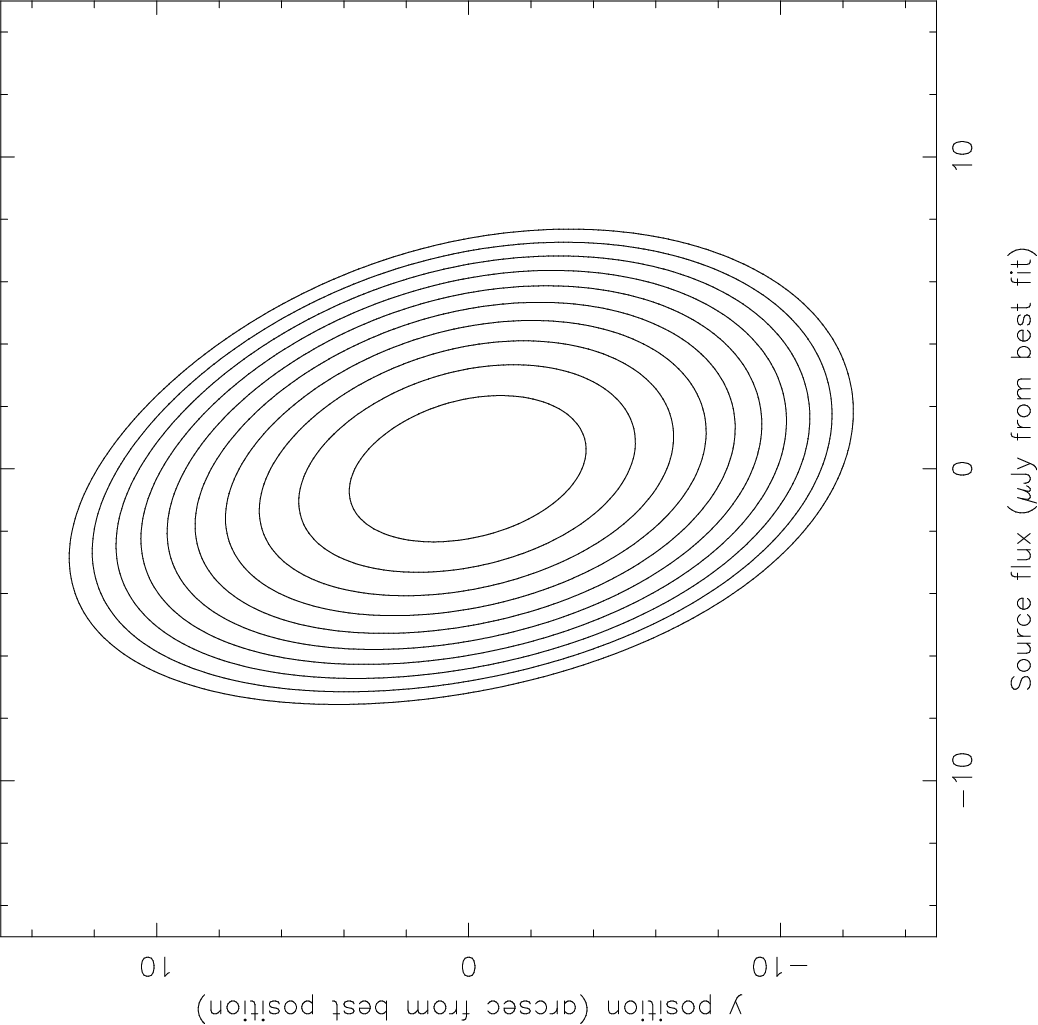}
\caption{$\chi^{2}$ contours for position for the
brightest source in the test field. The spacing between contour levels
in each figure is such that the reduced $\chi^2$ value increases by 1
between each contour.}
\end{centre}
\end{figure}

\textsc{FluxFitter} also reports error bounds. As
Figure~\ref{fig:chiplots} shows, the $\chi^{2}$ contours are elliptical
and orientated along the variable axes, which shows that the
parameters are independent. The error on each parameter is
calculated by finding the parameter values at which the reduced $\chi^{2}$
increases by 1. 

\begin{figure}
\begin{centre}
\epsfig{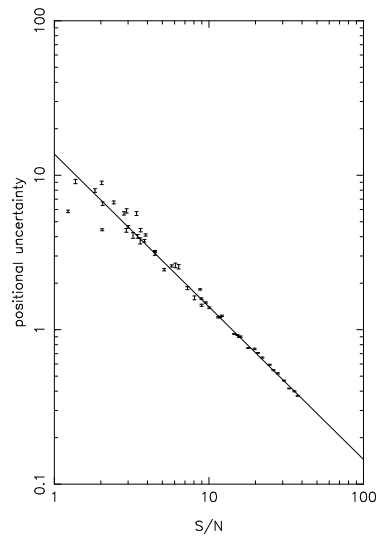}
\caption{Signal-to-noise ratio \textit{verses} the uncertainty (in arcseconds) of the
position. Fitting a straight line to this data gives
$\textrm{uncertainity}/\textrm{arcsec} = (13.7\pm0.5) (\mathrm{signal/noise})^{-0.99\pm0.01}$}
\label{fig:snplot}
\end{centre}
\end{figure}

The error-bound reporting was checked by simulating a point source
with differing signal-to-noise ratio.  500 observations of a single
point source were simulated; the signal-to-noise was kept constant for
groups of 10 simulations, and the position was held constant for all
the simulations.
The visibilities were simulated with Gaussian noise.  The
known position was then fed to \textsc{FluxFitter} as an initial
guess, and the best-fitting position and flux recorded. It was found
that the quoted error bar does enclose the position for 67\% of the
simulations. It was also found that the uncertainty of the position, that is the
size of $\sigma$, varies as the inverse of the signal-to-noise. This
result is shown in Figure~\ref{fig:snplot}. Note that this relation
holds down to very low signal-to-noise ratios.
The result is useful for determining whether a tentative source found
with the RT at low signal-to-noise has a position coincident with
higher significance data, for example from NVSS or POSS.

\subsubsection{Possible improvements}

Source recognition is clearly the biggest problem that still
remains. It is the only step that is still performed in the map plane
rather than aperture plane. There are computational issues involved
here: producing a map and identifying sources ``by hand'' is possible
and fairly cheap in computer time, but the noise in the map plane is
non-Gaussian. Minimising the misfit between the data and a given
number of sources is also cheap; but allowing the number of sources to
vary vastly increases the complexity of the problem and the time
required. It is possible that more advanced minimising techniques
such as simulated annealing (see \eg Press et al. 1993) or
using massive inference techniques will make this possible and robust
in the aperture plane.

\subsection{Comparison of results}

Tables~\ref{tab:clean2}, \ref{tab:Matrixpositions} and \ref{tab:flux2}
show the results of three different methods for source fitting. For
both the \clean\ and \textsc{FluxFitter} methods, the results with six
sources are considered.  The positions and fluxes put into the model
are shown in Table~\ref{tab:modelfluxes}. The resultant position and
depth of the \sz\ decrements after subtraction are shown in
Table~\ref{tab:szresults}. The parameters are all measured from dirty
maps, made with baselines shorter than 1~\kl. Note that the matrix
method has a less deep \sz\ flux density, and that the
\textsc{FluxFitter} and \clean\ method values are statistically
consistent with the expected value for this model cluster ($-660$
\microJy\ beam$^{-1}$).  Also, the positions of the \sz\ decrement are
fully consistent with the model as the beam size is around 180\arcsec.
It is not surprising that the flux of the \sz\ decrement from the
matrix method is less deep as the central 100~\microJy\ source has not
been subtracted. Also, the matrix method does an incomplete
subtraction of the sources it does find, resulting in more
contamination of the \sz\ signal. Note that this does not imply that
the matrix method will always will give a lower \sz\ decrement if the
source subtraction is incomplete.

\begin{table}
\caption{Parameters for the resulting \sz\ decrements from the simulated
data.}
\label{tab:szresults}
\begin{tabular}{lll}
Subtraction & \sz\ flux & position \\
method      & (\microJy\  beam$^{-1}$)   & (arcseconds)  \\
\clean & $-675 \pm 72$ &    $-$4, $-$2 \\
Matrix & $-500 \pm 90$ &    18, 0 \\
\textsc{FluxFitter} & $-700 \pm 65$ & 9, 4 \\
\end{tabular}
\end{table}

All three methods benefit from prior knowledge of the source
distribution on the sky. This can be estimated from looking at lower
frequency surveys such as NVSS or FIRST. Most falling-spectrum
sources, \ie with $\alpha > 0$ (where $S \propto v^{-\alpha}$ where
$v$ is the observing frequency and $\alpha$ is the spectral index)
will be detected in NVSS and or FIRST. However, as shown by
Cooray et al.\ \shortcite{cooray} in clusters and Taylor \shortcite{angela} generally, there are
rising spectrum sources, \ie with $\alpha < 0$ that are present at
15~GHz and not detected in NVSS and FIRST.

\section{Observations of Abell 611}

\subsection{\xray}

Abell 611 is a cluster at $z=0.288$ \cite{A611z} originally identified
by Abell \shortcite{Abell}. It has an 0.1--2.4 keV luminosity of
$8.63\times10^{44}$ W \cite{noras}, with a
temperature of $7.95^{+0.56}_{-0.52}\times 10^7$~K
\cite{WhiteTprofiles}. White derived this value from a 57-ks ASCA
exposure by considering both a single-phase and two-phase cooling
model. The temperature values found for the bulk of the gas are
statistically equivalent and a mass deposit rate of $0^{+177}_{-0}$
\msolar\ yr$^{-1}$ was found for the cooling model.

\begin{figure}
\begin{centre}
\epsfig{file=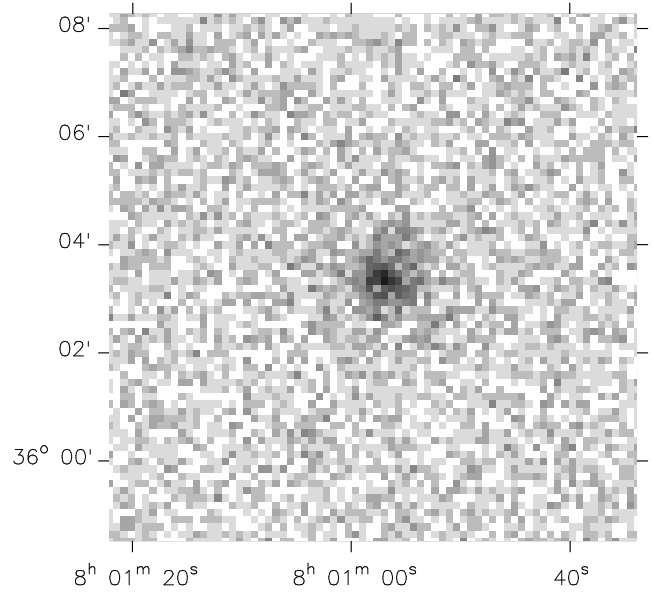}
\caption{\rosat\ HRI image of A611. The exposure time is
17 ks. The greyscale range is 0 to 32 counts.}
\label{fig:xrayimage}
\end{centre}
\end{figure}

The 17-ks \rosat\ HRI observation from April 1996 is shown in
Figure~\ref{fig:xrayimage}. The image contains two bright pixels,
which, on comparison with the POSS image, are coincident with a large
galaxy. These pixels were removed. We calculate an \xray\ emissivity
constant of 1.29$\times 10^{-69}$ counts s$^{-1}$ from 1~m$^{3}$ of
$7.95\times10^7$-K gas of electron
density 1~m$^{-3}$  at a luminosity distance of 1~Mpc, assuming a
metallicity of $0.21\pm0.07$ solar and an absorbing H column of
$4.88\times10^{24}$~m$^{-2}$. 

The best-fitting model parameters were $\beta = 0.59$, core radii of
26\arcsec\ and 24\arcsec\ with a position angle of the major axis of
101$^{\circ}$ and a central electron density of $n_0 =
11.6\times10^{3}$ m$^{-3}$ (assuming a core radius along the line of
sight of $25 = (24 \times 26)^{1/2}$ arcsec and \ho =
50~\hounit). There is a degeneracy between the core radii fitted and
$\beta$ but this has no significant effect on \ho\ (see Grainge
et al.\ 2001b and Jones et al.\ 2001). 

\subsection{RT observations}

A611 was observed for 16 sets of 12 hours between November 1994 and
January 1995 with the RT in configuration Cb. Flux and phase
calibration and overall data reduction strategy are described in
Grainge et al.\ \shortcite{0016}.  Three days of data -- taken in bad
weather -- were rejected after examining the 1-day maps and noise
levels. A map of the combined 13 days of data using baselines longer
than 1.5~\kl\ had a noise level of 70~\microJy\ beam$^{-1}$, and only
one source was visible, with flux 299~\microJy\ beam$^{-1}$ at
\ra8h0m57.1s \dec36d3m40s.  This source was removed with
\textsc{Uvsub}, using the flux and position from the dirty map. A
long-baseline map of the subtracted data was consistent with noise,
with no other sources in the field.

Table~\ref{tab:a611nvss} lists the sources found in the FIRST
catalogue around the pointing centre for A611. The NVSS catalogue
contains no sources in this region. Neither of the two FIRST sources
is detected at 15~GHz and the source that is present at 15~GHz is not
detected at lower frequencies. 

\begin{table}
\caption{The radio sources in the FIRST catalogue within 400\arcsec\
of the RT pointing centre for A611.}
\label{tab:a611nvss}
\begin{tabular}{lll}
 & & Flux (\milliJy\  beam$^{-1}$) \\
RA & Dec & Peak  \\
8 1 20.248 & 36 5 9.3 & $1.25 \pm 0.13$\\
8 0 54.948 & 36 9 6.1 & $1.10 \pm 0.13$\\
\end{tabular}
\end{table}

\begin{figure}
\begin{centre}
\epsfig{file=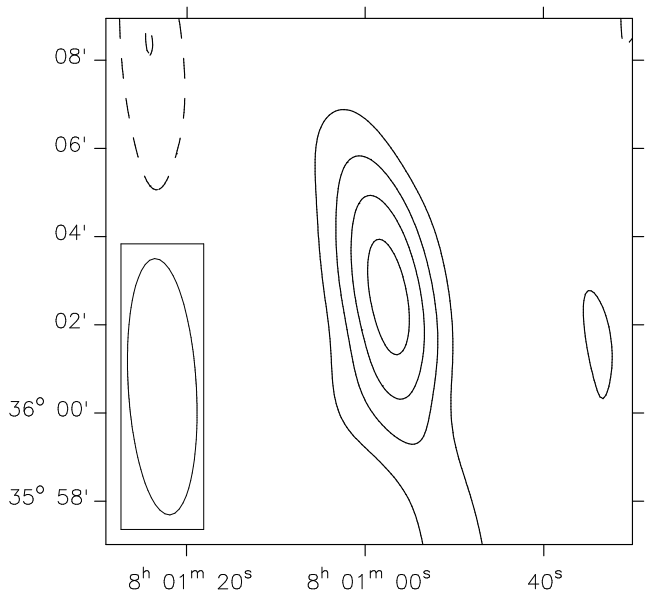}
\caption{The \sz\ effect in A611. The contour levels are $-$480,
$-$360, $-$240 and $-$120~\microJy\ beam$^{-1}$\ (solid) and
120~\microJy\ beam$^{-1}$\ (dashed). The map has been
\textsc{Clean}ed; the restoring beam, which is 92\arcsec\ by
350\arcsec\ FWHM at a position angle 3.4$^{\circ}$, is shown in the
bottom left.}
\label{fig:a611short}
\end{centre}
\end{figure}

\textsc{FluxFitter} was then run using the \xray\ data to provide a
model of the \sz\ decrement and using all the baselines. Again, the
initial guess was defined by the 1.5-\kl -only fitting. The source was
found to be at \ra8h0m57.1s $\pm 0.9$ \dec36d3m35s0 $\pm 8$ with a
flux of $188 \pm 65$~\microJy\  beam$^{-1}$, which is lower than the long-baseline
only values. As the angular size of A611 is small, it is likely that
the \sz\ signal was contaminating the ``long''-baseline
map. Note that this source is not detected in the FIRST survey, and so
in this case prior knowledge from a lower frequency survey has not helped.
Figure~\ref{fig:a611short} shows a \clean ed map of baselines
shorter than 1~\kl\ after this source has been subtracted. The
decrement (as measured from the map) is $-540 \pm 125$~\microJy\ beam$^{-1}$ at
\ra8h0m57.3s \dec36d2m38s. This location is 3\arcsec\ in RA and
36\arcsec\ in Dec away from the \xray\ cluster location. Considering
the \clean\ beam used is 92\arcsec $\times$ 350\arcsec, this is a good
positional agreement between the \xray\ and \sz\ observations.  The
slight extension to the south is not significant.

\subsection{\ho\ determination}

\begin{figure}
\begin{centre}
\epsfig{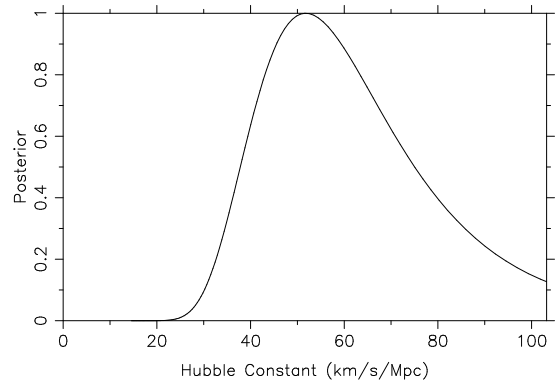}
\caption{Likelihood plot for different \ho\ values from fitting to the
source-subtracted \sz\ data from A611.}
\label{fig:a611likes}
\end{centre}
\end{figure}

From the source subtracted dataset and the \xray\ parameters of the
cluster, it is possible to estimate \ho, as described in Grainge et
al.\ \shortcite{1413}.  The likelihoods for different \ho\ values are
calculated and are shown in Figure~\ref{fig:a611likes}.  The best-fit
\ho\ for A611 is then $52^{+22}_{-14}$~\hounit.  The error quoted is
due to the noise in the \sz\ measurement, and does not include any of
the other sources of error in the determination. The additional
sources of error are described fully in Grainge et al.\
\shortcite{1413}.  For A611, the error from the \sz\ measurement is by
far the most important, and the final \ho\ value is
$52^{+24}_{-16}$~\hounit\ for an Einstein-de-Sitter world model. Assuming a
world model with $\Omega_{\Lambda}=0.7$, $\Omega_{\rm M}=0.3$,
\ho=$59^{+27}_{-18}$~\hounit\ from this cluster.

\section{Conclusions}

The problem of radio source contamination in interferometric \sz\
observations and methods to remove it have been
investigated, demonstrating the following.

\begin{enumerate}
\renewcommand{\theenumi}{(\arabic{enumi})}
\item
The non-linear \clean\ method can work well, but does not use all the
available information and can over-complicate the problem.

\item
The matrix method, though linear, fails in typical situations such as the one
simulated here. The failure is mainly due to the high sidelobes from the
Ryle Telescope; they make it difficult to determine accurate positions,
and so fluxes, for sources close to each other on a map. 

\item 
\textsc{FluxFitter} uses all the available information and produces
the simplest model. It solves simultaneously for sources and \sz\
decrement, and it works with with the visibilities, where noise is
known to be Gaussian, rather than in the map plane where the noise is
correlated.

\item
All three techniques suffer from the problem of source
identification, which is currently performed in the image plane where
the noise characteristics are complex. Source identification can be
aided by prior information, for example from lower-frequency surveys. 

\item
The positional uncertainty, as determined in the aperture plane, is
found to vary as $\textrm{uncertainity}/\textrm{arcsec} \propto
(\mathrm{signal/noise})^{-0.99\pm0.01}$ even at signal-to-noise ratios
below nominal detection limits. The constant of proportionality will
be a function of the interferometer used. 

\item
Observations of the cluster A611 with the Ryle Telescope give a
4.3-$\sigma$ detection of an \sz\ decrement, and combination with \xray\
data gives and estimate of \ho = $52^{+24}_{-16}$~\hounit, assuming an
Einstein-de-Sitter cosmology, and $59^{+27}_{-18}$~\hounit\ using
$\Omega_{\Lambda}=0.7$ and $\Omega_{\rm M}=0.3$.

\end{enumerate}

\section{Acknowledgements}

We thank the staff of the Cavendish Astrophysics group who ensure the
continued operation of the Ryle Telescope. Operation of the RT is
funded by PPARC.  WFG acknowledges support from a PPARC
studentship. We have made use of the ROSAT Data Archive of the
Max-Planck-Institut f\"ur extraterrestrische Physik (MPE) at Garching,
Germany.

\end{document}